\documentclass[12pt,preprint]{emulateapj}

\shorttitle{SN2009bb Explosion Environment}
\shortauthors{Levesque et al.}
\slugcomment{DRAFT \today}

\begin{document}

\title{The High-Metallicity Explosion Environment of the Relativistic Supernova 2009bb$^1$}
\author{E. M. Levesque\footnotemark[2,]\footnotemark[3], A. M. Soderberg\footnotemark[3,]\footnotemark[4], R. J. Foley\footnotemark[3,]\footnotemark[5], E. Berger\footnotemark[3], L. J. Kewley\footnotemark[2], S. Chakraborti\footnotemark[6], A. Ray\footnotemark[6], M. A. P. Torres\footnotemark[3], P. Challis\footnotemark[3], R. P. Kirshner\footnotemark[3], S. D. Barthelmy\footnotemark[7], M. F. Bietenholz\footnotemark[8,]\footnotemark[9], P. Chandra\footnotemark[10], V. Chaplin\footnotemark[11], R. A. Chevalier\footnotemark[12], N. Chugai\footnotemark[13], V. Connaughton\footnotemark[11], A. Copete\footnotemark[3], O. Fox\footnotemark[12], C. Fransson\footnotemark[14], J. E. Grindlay\footnotemark[3], M. A. Hamuy\footnotemark[15], P. A. Milne\footnotemark[16], G. Pignata\footnotemark[15,17], M. D. Stritzinger\footnotemark[18,]\footnotemark[19], M. H. Wieringa\footnotemark[20]}

\email{emsque@ifa.hawaii.edu}

\footnotetext[1]{This paper is based on data gathered with the 6.5 m Magellan telescopes located at Las Campanas, Chile.}
\footnotetext[2]{Institute for Astronomy, University of Hawaii, 2680 Woodlawn Dr., Honolulu, HI 96822}
\footnotetext[3]{Harvard-Smithsonian Center for Astrophysics, 60 Garden St., MS-51, Cambridge, MA 02138}
\footnotetext[4]{Hubble Fellow}
\footnotetext[5]{Clay Fellow}
\footnotetext[6]{Tata Institute of Fundamental Research, Mumbai 400 005, India}
\footnotetext[7]{NASA Goddard Space Flight Center, Greenbelt, MD 20771, USA}
\footnotetext[8]{Department of Physics and Astronomy, York University, Toronto, ON M3J 1P3, Canada}
\footnotetext[9]{Hartebeestehoek Radio Observatory, PO Box 443, Krugersdorp, 1740, South Africa}
\footnotetext[10]{Royal Military College of Canada, Kingston, ON Canada}
\footnotetext[11]{University of Alabama, Huntsville, AL, USA}
\footnotetext[12]{University of Virginia, Department of Astronomy, PO Box 400325, Charlottesville, VA 22904, USA}
\footnotetext[13]{Institute of Astronomy, RAS, Pyatnitskaya 48, Moscow 119017, Russia}
\footnotetext[14]{Department of Astronomy, Stockholm University, AlbaNova, SE-106 91 Stockholm, Sweden}
\footnotetext[15]{Departamento de Astronomia, Universidad de Chile, Casilla 36-D, Santiago, Chile}
\footnotetext[16]{Steward Observatory, University of Arizona, 933 North Cherry Avenue, Tucson, AZ 85721, USA}
\footnotetext[17]{Departamento de Ciencias Fisicas, Universidad Andres Bello, Avda. Republica 252, Santiago, Chile}
\footnotetext[18]{Las Campanas Observatory, Carnegie Observatories, Casilla 601, La Serena, Chile}
\footnotetext[19]{Dark Cosmology Centre, Niels Bohr Institute, University of Copenhagen, Copenhagen, Denmark}
\footnotetext[20]{Australia Telescope National Facility, CSIRO, Epping 2121, Australia}

\begin{abstract}
We investigate the environment of the nearby ($d \approx 40$ Mpc) broad-lined Type Ic supernova SN 2009bb. This event was observed to produce a relativistic outflow likely powered by a central accreting compact object. While such a phenomenon was previously observed only in long-duration gamma-ray bursts (LGRBs), no LGRB was detected in association with SN 2009bb. Using an optical spectrum of the SN 2009bb explosion site, we determine a variety of ISM properties for the host environment, including metallicity, young stellar population age, and star formation rate. We compare the SN explosion site properties to observations of LGRB and broad-lined SN Ic host environments on optical emission line ratio diagnostic diagrams. Based on these analyses, we find that the SN 2009bb explosion site has a metallicity between $1.7Z_{\odot}$ and $3.5Z_{\odot}$, in agreement with other broad-lined SN Ic host environments and at odds with the low-redshift LGRB host environments and recently proposed maximum metallicity limits for relativistic explosions. We consider the implications of these findings and the impact that SN 2009bb's unusual explosive properties and environment have on our understanding of the key physical ingredient that enables some SNe to produce a relativistic outflow.
\end{abstract}

\section{Introduction}
Relativistic supernovae (SNe) mark the explosive deaths of massive stars, and until recently were discovered exclusively through their association with long-duration gamma-ray bursts (LGRBs). Thanks to the discovery of several LGRBs at $z \lesssim 0.3$, we now know that these are Type Ic SNe (SNe Ic) with broad absorption lines (hereafter ``broad-lined"; see Woosley \& Bloom 2006 and references therein). These LGRB-associated SNe are distinguished from ordinary SN Ic explosions by the production of a relativistic outflow powered by a central engine (an accreting black hole or neutron star) which gives rise to the gamma-ray emission and a non-thermal afterglow (Piran 1999).  A dichotomy is also indicated by their relative rates, with just $0.1-1\%$ of Ic SNe giving rise to an LGRB after accounting for collimation of the ejecta (Soderberg et al.\ 2006a, 2006b, Guetta \& Della Valle 2007). While these relativistic ejecta are typically manifested in the form of LGRBs, this is not an exclusive association (see, for example, X-ray flashes; Heise et al.\ 2001), and the gamma-ray emission does not always dominate the total relativistic yield. In the case of GRB 980425 and the associated SN 1998bw, the total energy released high-energy emission was dwarfed by the kinetic energy of the blast-wave by a factor of 100, as inferred from radio observations (Kulkarni et al.\ 1998).

While there is growing evidence that LGRBs and SNe Ic share Wolf-Rayet (WR) progenitor stars (e.g., Woosley et al.\ 2002), the critical physical ingredient that enables only a small fraction to explode relativistically remains unknown.  Numerical simulations of the explosions indicate that high angular momentum may be the critical physical parameter of LGRB progenitors (MacFayden et al.\ 2001, Dessart et al.\ 2008). Since the metallicity-dependent line-driven winds of WR stars serve to strip away angular momentum (Woosley \& Heger 2006), low metallicities of $Z\lesssim 0.3~Z_{\odot}$ have been proposed for LGRB progenitors as a means of reducing the line-driven mass loss rate and sustaining this fast rotation - a relation $\dot{M}\propto Z^{0.86}$ has been calculated for WN-type WR stars (Vink \& de Koter 2005).

Observations of nearby ($z\lesssim 0.3$) LGRB host galaxies offer some support for this theoretical prediction, with typical metallicities estimated at $\lesssim 0.5~Z_{\odot}$ (Levesque et al.\ 2009a), lower on average than those inferred for the explosion sites of local broad-lined SNe Ic (Modjaz et al.\ 2008).  In such work, the inferred environmental metallicities are considered to be representative of the natal properties of the progenitor stars (S\'{i}mon-D\'{i}az et al.\ 2006, Hunter et al.\ 2007). However, the nearby LGRB sample remains small and observations of additional central engine-driven explosions are required to test the dependence of explosion properties on the inferred progenitor metallicity.

SN 2009bb marks the first relativistic SN discovered without a gamma-ray trigger. It was first detected on 2009 Mar 21.11 (UT) in the nearby ($d \approx 40$ Mpc) face-on spiral galaxy NGC 3278 (Pignata et al.\ 2009). Stritzinger et al.\ (2009) examined an optical spectrum of the event and classified it as a broad-lined SN Ic. Radio observations revealed that SN 2009bb also produced a relativistic explosion likely powered by a central engine. Such outflows have previously only been observed in LGRBs; however, satellites reveal no detected LGRB in association with SN 2009bb (Soderberg et al.\ 2009).

Here we present observations of the explosion site of the relativistic SN 2009bb (Section 2). We use these observations to determine a variety of ISM properties, including metallicity, young stellar population age, and star formation rate (Section 3). We plot SN 2009bb on two optical emission line ratio diagnostic diagrams, allowing a comparison between its ISM environment, stellar population synthesis and photoionization models, and other nearby galaxies (Section 4). We discuss our findings and their implications for SN 2009bb and our understanding of engine-driven relativistic explosions (Section 5).

\section{Data Acquisition}
\subsection{Observations}
We observed the SN 2009bb explosion environment with the MagE spectrograph (Marshall et al.\ 2008) mounted on the Magellan/Clay 6.5m telescope on Apr 26.1 UT with a 1.0 arcsec slit for 1800 sec in good conditions, obtaining data on a $\sim$1 arcsec$^2$ (190 $\times$ 190 pc) region centered on the explosion site (Figure 1). The extraction box was 1" $\times$ 0.75". CCD processing and spectral extraction were carried out with standard IRAF\footnotemark[2] \footnotetext[2]{IRAF is distributed by NOAO, which is operated by AURA, Inc., under cooperative agreement with the NSF.} packages. The data were extracted using an optimal extraction algorithm. Low-order polynomial fits to calibration-lamp spectra were used to establish the wavelength scale, and small adjustments derived from night-sky lines in the object frames were applied. The sky was subtracted from the images using the method described by Kelson (2003). We employed our own IDL routines (see Foley et al.\ 2009 and references therein) to flux calibrate the data and remove telluric lines using the well exposed continuum of the spectrophotometric standard Hiltner 600.

Strong Balmer series emission lines from the underlying star-forming region dominate the spectrum. We also clearly identify [OII]$\lambda$3727, [OIII]$\lambda\lambda$4959,5007, [NII]$\lambda\lambda$6549,6583, and [SII]$\lambda\lambda$6717,6730 emission lines, as well as the Na I D feature in absorption for the Milky Way as well as the SN 2009bb host. To isolate the host galaxy emission at the explosion site, we subtract a high order polynomial fit to the broad SN spectral features. The explosion site spectrum both before and after the subtraction of the SN contribution is shown in Figure 2.

\section{Physical Properties of the SN 2009bb Environment}
\subsection{Extinction}
We determine E($B-V$) based on the observed H$\alpha$ and H$\beta$ line fluxes and the reddening law from Cardelli et al.\ (1981) with the standard total-to-selective extinction ratio $R_V = 3.1$. We assume a Balmer decrement of H$\alpha$/H$\beta$ = 2.87 (following Osterbrock 1987 for case B recombination) and the wavelength-dependent constant k(H$\alpha$) = 2.535 (Cardelli et al.\ 1981). We find a total line-of-sight E($B-V$) = 0.48 mag for the explosion site, in excess of the Galactic E($B-V$) $\approx$ 0.098 in the direction of SN 2009bb (Schlegel et al.\ 1998), and use this value to correct for extinction effects in the observed fluxes measured in the emission line spectrum. This value is included in Table 1.

\subsection{Metallicity}
The metallicity diagnostics presented in Kewley \& Dopita (2002; see also Kewley \& Ellison 2008) determine metallicity based on strong optical emission line ratios through equations derived from photoionization models. We adopt the Kewley \& Dopita (2002) polynomial relation between the [NII]/[OII] ratio and metallicity, and find log(O/H) + 12 = 9.0 $\pm$ 0.1, or $Z \approx 2Z_{\odot}$ (where the solar metallicity is log(O/H) + 12 = 8.7, following Asplund et al.\ 2005). By comparison, the Kewley \& Dopita (2002) metallicites found for nearby ($z < 0.3$) LGRB host galaxies range from 8.1 $<$ log(O/H) + 12 $<$ 8.4, while the Kewley \& Dopita (2002) metallicites of the SN Ic hosts range from 8.6 $<$ log(O/H) + 12 $<$ 9.0 (Modjaz et al.\ 2008, Levesque et al.\ 2009a). It should be noted that, while these diagnostics are self-consistent, the metallicities determined by the Kewley \& Dopita (2002) cannot be considered absolute.

\subsection{Young Stellar Population Age}
The equivalent width of the H$\beta$ emission line is strongly dependent on the evolution of the HII region (Copetti et al.\ 1986). Levesque et al.\ (2009a) derive metallicity-dependent equations that relate H$\beta$ equivalent width and age, based on the evolutionary synthesis models of Schaerer \& Vacca (1998). Using these equations, we find a young stellar population age of 4.5 $\pm$ 0.5 Myr at the site of SN2009bb (Table 1), in good agreement with the 3-5 Myr age range expected for Wolf-Rayet stars, particularly at solar and super-solar metallicities (e.g., Schaerer et al.\ 1993 and references therein). 

\subsection{Star Formation Rates}
The host galaxy of SN2009bb, NGC 3278, is a star-forming galaxy with a diameter of $\sim$1 arcmin. Broadband optical data indicate an integrated luminosity of $M_B = -19.98 \pm 0.02$ mag (Lauberts \& Valentijn 1989), comparable to those of other nearby broad-lined SNe Ic (Prieto et al.\ 2008, Modjaz et al.\ 2008). Mid-IR and radio observations reveal the host galaxy to be luminous at longer wavelengths, reminiscent of starburst galaxies with elevated star-formation rates. Observations of the host galaxy from the literature with the Infrared Astronomy Satellite (IRAS; Sanders et al.\ 2003) and the Very Large Array (VLA; Mauch \& Sadler 2007), combined with our own measurements of the 617 Mhz integrated flux the Giant Metrewave Radio Telescope (GMRT; see Soderberg et al.\ Suppl. Info), indicate integrated luminosities of $L_{IR} \approx 1.2 \times 10^{44}$ erg s$^{-1}$ (8 - 1000 $\mu$m), $L_{1.4GHz} \approx 1.3 \times 10^{37}$ erg s$^{-1}$, and $L_{617MHz} \approx 1.3 \times 10^{38}$ erg s$^{-1}$, respectively. Adopting the SFR determinations of Kennicutt (1998) and Yun \& Carilli (2002), we estimate an integrated SFR $\approx 5-7 M_{\odot}$ yr$^{-1}$. We construct radio-to-optical galaxy templates from Silva et al. (1998) and Yun \& Carilli (2002), and compare these with the observed spectral energy distribution (SED) for NGC3278. As shown in Figure 3, the broadband SED of the host galaxy is most consistent with an Sc spiral, indicative of a stronger SFR than a typical Sa galaxy and a lower SFR than the extreme case of Arp 220.

From our emission line spectrum, we calculate the SFR within a $\sim$1 arcsec$^2$ region local to the SN based on the H$\alpha$ luminosity relation of Kennicutt (1998), and find 0.003 M$_{\odot}$ yr$^{-1}$. Soderberg et al.\ (2009) find that SN 2009bb coincides with the brightest and bluest region of the galaxy, which is indicative of star-forming activity and consistent with other Ic SNe and LGRBs (Kelly et al.\ 2008).

\section{Comparison with Nearby ($z < 0.3$) Galaxy Samples}
In Figure 4 we plot the SN 2009bb explosion site properties on two optical emission line diagnostics diagrams. These diagrams allow us to directly compare this ISM environment to other nearby ($z < 0.3$) galaxies. Our comparison populations include (1) a sample of local ($z \lesssim 0.14$) broad-lined SNe Ic global host galaxy spectra and explosion site spectra (Modjaz et al.\ 2008); (2) nearby ($z < 0.3$) LGRB host galaxies published in Levesque et al.\ (2009a), including observations of the GRB 980425 explosion site from Christensen et al.\ (2008); and (3) 60,920 local ($z < 0.1$) star-forming galaxies from SDSS (Kewley et al.\ 2006).

On these grids we also include the stellar population synthesis and photoionization model grids of Levesque et al.\ (2009b). We plot these grids using lines of constant metallicity to illustrate the metallicity of the SN 2009bb explosion site. In this comparison we adopt models which assume a zero-age instantaneous burst star formation history, an electron density $n_e = 100$ cm $^{-3}$, and an ionization parameter $-3.5 < \mathcal{U} < -1.9$ (Levesque et al.\ 2009b and references therein).

We first compare the SN2009bb explosion environment and comparison samples on the [NII]/H$\alpha$ vs. [OIII]/H$\beta$ grid of Baldwin et al.\ (1981) in Figure 4 (left). [NII]/H$\alpha$ correlates strongly with both metallicity and ionization parameter (Kewley \& Dopita 2002), while [OIII]/H$\beta$ is primarily a measure of ionization parameter with a double-valued dependence on metallicity (Baldwin et al.\ 1981). On this diagram, we see that the SN2009bb explosion site environment is similar to the host galaxies and explosion site environments of SNe Ic, as well as the sample of SDSS star-forming galaxies. By contrast, the SN 2009bb environment is distinct from LGRB host galaxies, which fall in a lower-metallicity region of the diagnostic diagram according to the model grids. A comparison with the model grid also illustrates that the SN 2009bb explosion site has a super-solar metallicity, falling between the $1.7Z_{\odot}$ and $3.5Z_{\odot}$ models.

The [NII]/[OII] vs. [OIII]/[OII] diagnostic diagram of Dopita et al.\ (2000) (Figure 4, right) is more sensitive to extinction, but removes the degeneracy present in the Baldwin et al.\ (1981) diagram between metallicity and ionization parameter. [NII]/[OII] is strongly dependent on metallicity with a minimal dependence on ionization parameter (Dopita et al.\ 2000), while [OIII]/[OII] is mainly reliant on ionization parameter, with a non-degenerate dependence on metallicity (Dopita et al.\ 2000, Kewley \& Dopita 2002). We again see that the position of the SN2009bb explosion environment is similar to the SNe Ic host environments and the SDSS galaxies, while the LGRB host environments occupy a distinct and lower-metallicity region of the diagnostic diagram. Comparing the galaxy samples to the model grids on this diagnostic diagram, we again see that the SN2009bb explosion environment falls on the super-solar lines of constant metallicity on the grids, with $1.7 - 3.5Z_{\odot}$ models and at the high end of the SNe Ic distribution. Four of the five LGRB host environments, on the other hand, fall on the lowest-metallicity lines of the grid. The exception to this is the host environment of GRB 031203, which appears from the diagram to have a super-solar metallicity ($\approx1.7Z_{\odot}$).\footnotemark[3] \footnotetext[3]{Levesque et al.\ (2009a) suggest that this host galaxy's spectrum may show evidence of AGN activity (which would hinder drawing any robust conclusions about the ISM environment based on the emission line diagnostic diagrams), and find a much lower metallicity based on the Kewley \& Dopita (2002) diagnostic. The host environment of SN 2009bb does not show any similar signs of AGN activity.}

\section{Discussion}
SN 2009bb marks the first discovery of a relativistic SN in a high-metallicity environment; previously, such observations have been restricted to SNe with accompanying GRBs in low-metallicity environments. We find that the SN 2009bb explosion site has a very high metallicity of $Z \approx 2Z_{\odot}$ that differs from those of LGRB host galaxies (Levesque et al.\ 2009) but is in good agreement with those of the broad-lined SNe Ic host galaxies (Modjaz et al.\ 2008). The age of the young stellar population, the star-formation rate of the host galaxy, and the mass loss rate of the progenitor (Soderberg et al.\ 2009) are all consistent with {\it both} broad-lined SNe Ic and LGRB host environments.

The high metallicity of the SN 2009bb explosion environment is at odds with several recent studies of LGRB host environments, which propose metallicity cut-offs of log(O/H) + 12 $\approx$ 8.5-8.7 for environments that can produce such explosions (Wolf \& Podsiadlowski 2007, Modjaz et al.\ 2008, Kocevski et al.\ 2009). The observation of a relativistic SN in a high-metallicity environment raises several critical questions that must be considered in future studies of these events: 

1) {\it Does SN 2009bb come from the same progenitor population as nearby LGRBs?} Addressing this question requires future host environment observations of more nearby engine-driven relativistic SNe, both {\it with} and {\it without} gamma-ray triggers. With the Panoramic Survey Telescope And Rapid Response System (Pan-STARRS; Kaiser et al.\ 2002) and the Palomar Transient Factory (PTF; Law et al.\ 2009) surveys now online, we expect to discover Ic supernovae and identify a relativistic subset with the Expanded VLA at a rate of $\sim$ 1 yr$^{-1}$ (Soderberg et al.\ 2009). This will allow us to pinpoint and observe host environments of relativistic events that may have otherwise gone undetected, and to assess the continuum of events with and without gamma-ray emission.

2) {\it What can the host environments tell us about the nature of relativistic explosions?} It is our expectation that the host environment of any event with a massive star progenitor should be indicative of the progenitor metallicity itself (e.g., Sim\'{o}n-D\'{i}az et al.\ 2006, Hunter et al.\ 2007). However, we currently have a limited understanding of the correlation between host environments, progenitor properties, and the nature of these powerful stellar explosions. Stanek et al.\ (2006) propose a correlation between metallicity and the isotropic gamma-ray energy release for the nearby ($z < 0.3$) LGRBs, finding that energy decreases steeply with increasing metallicity; however, the correct approach should be to consider the total relativistic energy output, since the blast-wave kinetic energy can dominate (Kulkarni et al.\ 1998). Improving our understanding of the relationship between host environments and explosive properties requires a rigorous comparison of these parameters for both LGRBs  and engine-driven relativistic supernovae such as SN 2009bb; this will allow us to investigate any potential correlations between metallicity and the energetic properties that remain the defining signatures of these explosions.

3) {\it How can an engine-driven relativistic explosion be produced by a high-metallicity progenitor?} The high metallicity of the SN environment combined with the lack of any detected accompanying LGRB sets SN 2009bb as a highly unusual engine-driven relativistic explosion. Several recent studies suggest that such explosions, which up until now have only been observed in connection with LGRBs, occur in metal-poor environments (Stanek et al.\ 2006, Modjaz et al.\ 2008, Kocevski et al.\ 2009, Levesque et al.\ 2009a). This correlation is supported by current theoretical models (Woosley \& Heger 2006, Yoon et al.\ 2006). However, SN 2009bb demonstrates that engine-driven relativistic explosions can also occur at high metallicity. More recent work on the evolution of the progenitors has also suggested that low metallicity might not be required, and in fact may even inhibit the production of a relativistic explosion (Dessart et al.\ 2008). Additionally, observations show that the expected progenitors of these events -- WC and WO-type Wolf-Rayet stars -- become increasingly rare in lower-metallicity environments (Massey 2003). To properly understand the origin of SN 2009bb and similar events, we require a more detailed investigation of the specific physical properties required to produce such phenomena in massive stars.

SN 2009bb challenges our current understanding of core-collapse supernovae and LGRBs. Our observations of an engine-driven relativistic explosion in a high-metallicity environment demonstrate that there is no maximum metallicity cut-off for the production of relativistic SNe. Discovery of this event in the absence of any detected accompanying LGRB (Soderberg et al.\ 2009) suggests that future searches for relativistic SNe should pursue identification based on joint optical and radio observations rather than being limited only to those accompanied by a gamma-ray trigger. Such surveys offer the most robust means of generating a complete sample of engine-driven relativistic SNe and probing the explosive properties and host environments of these enigmatic events.

We gratefully acknowledge the hospitality and assistance of the staff at Las Campanas Observatory in Chile. GMRT is run by the National Centre for Radio Astrophysics of the Tata Institute of Fundamental Research.This work made use of the Central Bureau for Astronomical Telegrams. We also include data from the Two Micron All Sky Survey (2MASS), which is a joint project of the University of Massachusetts and the Infrared Processes and Analysis Center, California Institute of Technology, funded by the National Aeronautics and Space Administration and the National Science Foundation. EL's participation was made possible in part by a Ford Foundation Predoctoral Fellowship. LK and EL gratefully acknowledge support by NSF EARLY CAREER AWARD AST07-48559. AS acknowledges support by NASA through a Hubble Fellowship grant. AR and SC's research is part of the 11th Five Year Plan Project No: 11P-409 at TIFR. RC and RK acknowledge support through NSF grants. GP ackowledges partial support from the Millennium Center for Supernova Science through grant P06-045-F funded by ``Programa Bicentenario de Ciencia y Tecnolog\'ia de CONICYT'' and ``Programa Iniciativa Cient\'ifica Milenio de MIDEPLAN'', ComiteMixto ESO-GOBIERNO DE CHILE, and the Center of Excellence in Astrophysics and Associated Technologies (PFB 06).

\begin{figure}
\epsscale{0.7}
\plotone{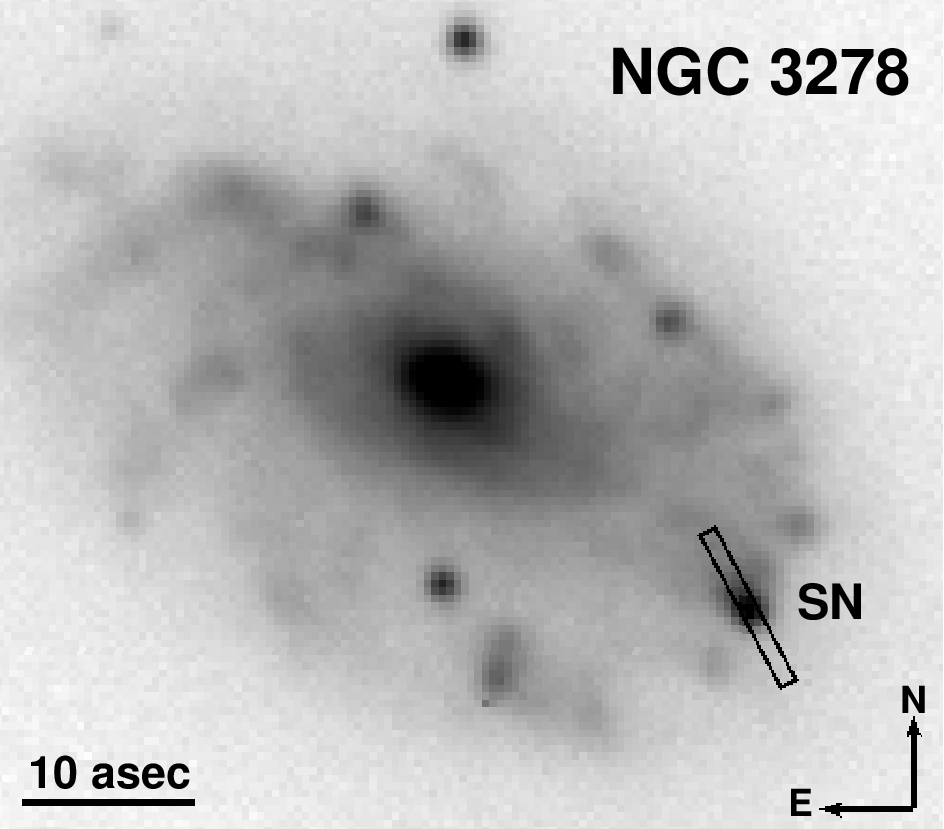}
\caption{An image of the host galaxy, NGC 3278, and SN\,2009bb was obtained with the SWOPE telescope at Las Campanas Observatory shortly after explosion (see Pignata et al.\ 2009 for details).  Our MagE spectrum was obtained with a 10" x 1" slit centered on the SN at a position angle of 28 degrees.}
\end{figure}

\begin{figure}
\epsscale{0.95}
\plotone{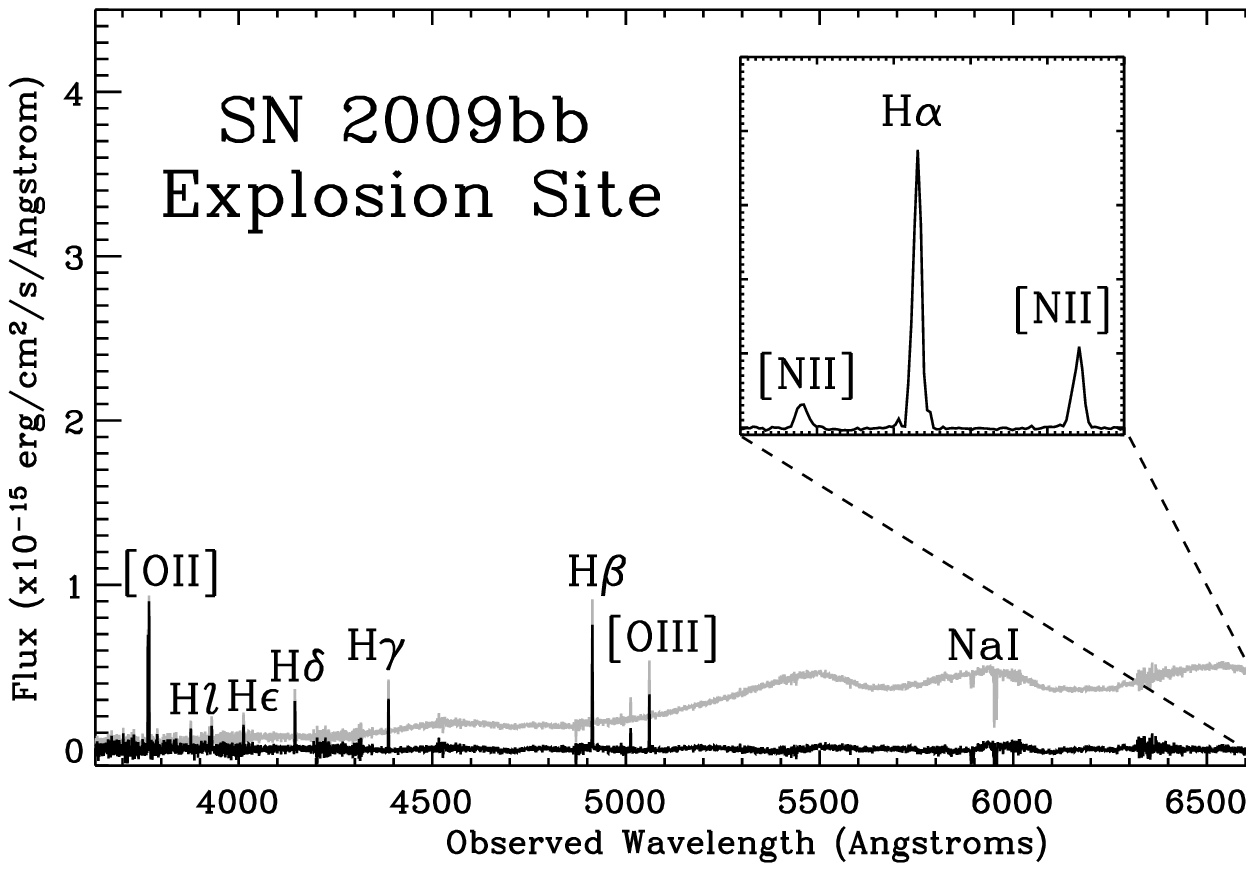}
\caption{Emission line spectrum taken at the SN2009bb explosion site both before (grey) and after (black) subtraction of the supernova contribution.}
\end{figure}

\begin{figure}
\epsscale{0.7}
\plotone{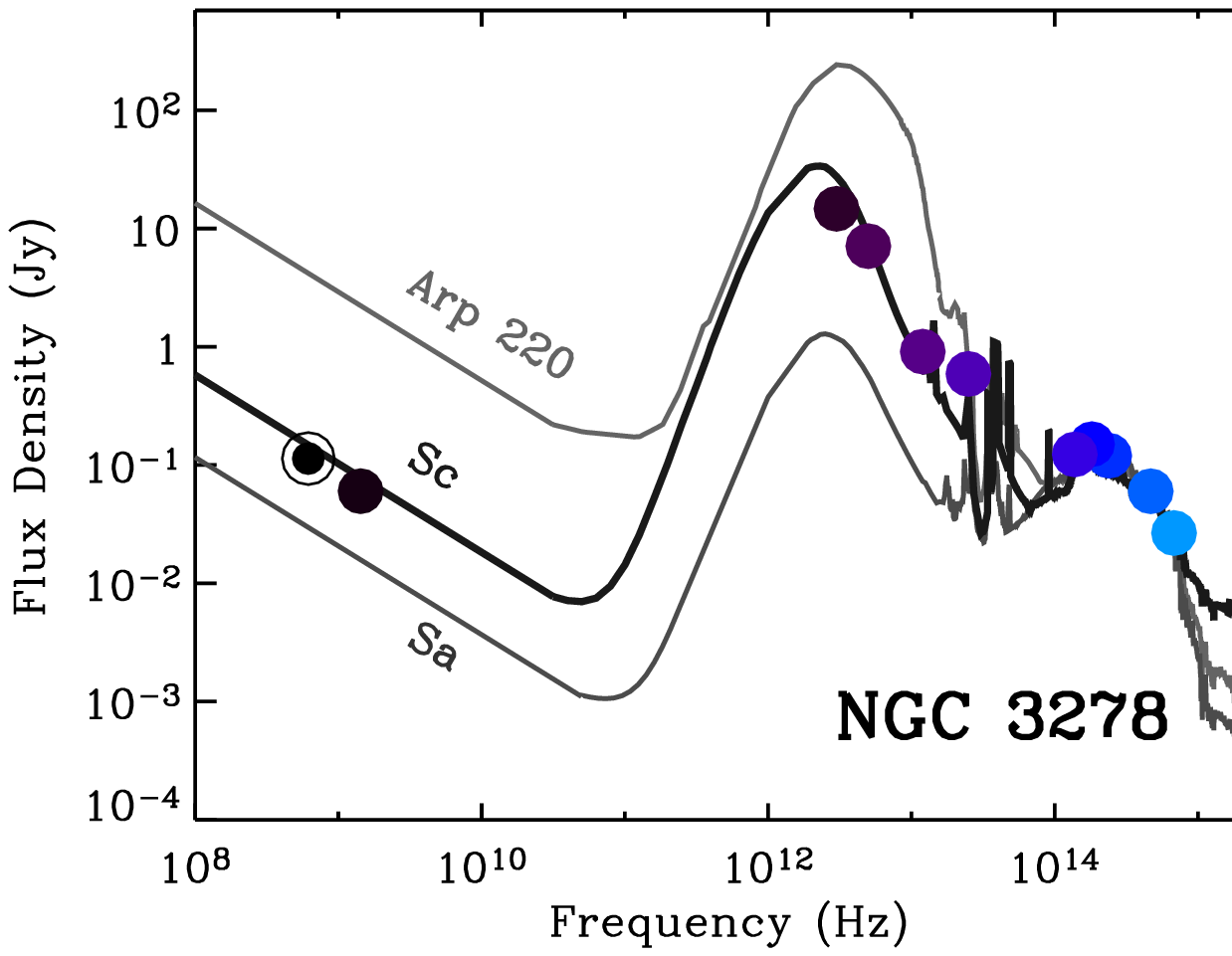}
\caption{The spectral energy distribution is shown for the host galaxy, NGC 3278, from radio to optical frequencies. We compiled integrated broadband flux densities for NGC 3278 extending from the optical (Lauberts \& Valentijn 1989) and near-IR bands (Two Micron All Sky Survey; 2MASS), to the mid-IR (IRAS; Sanders et al.\ 2003) radio wavelengths (VLA; Mauch \& Sadler 2007) and combined them with our 617 MHz measurements from the GMRT (this work). The galaxy is strongly star-forming (SFR $\approx 5 - 7$ M$_{\odot}$ yr$^{-1}$) as evidenced by the bright mid-IR emission. In comparison with standard galaxy templates (gray lines, from Silva et al.\ 1998), the spectrum of the host galaxy is most consistent with the broadband spectrum of an Sc galaxy with a star-formation which is elevated compared to a standard Sa spiral galaxy, but not as high as Arp 220.}
\end{figure}

\begin{figure}
\epsscale{0.5}
\plotone{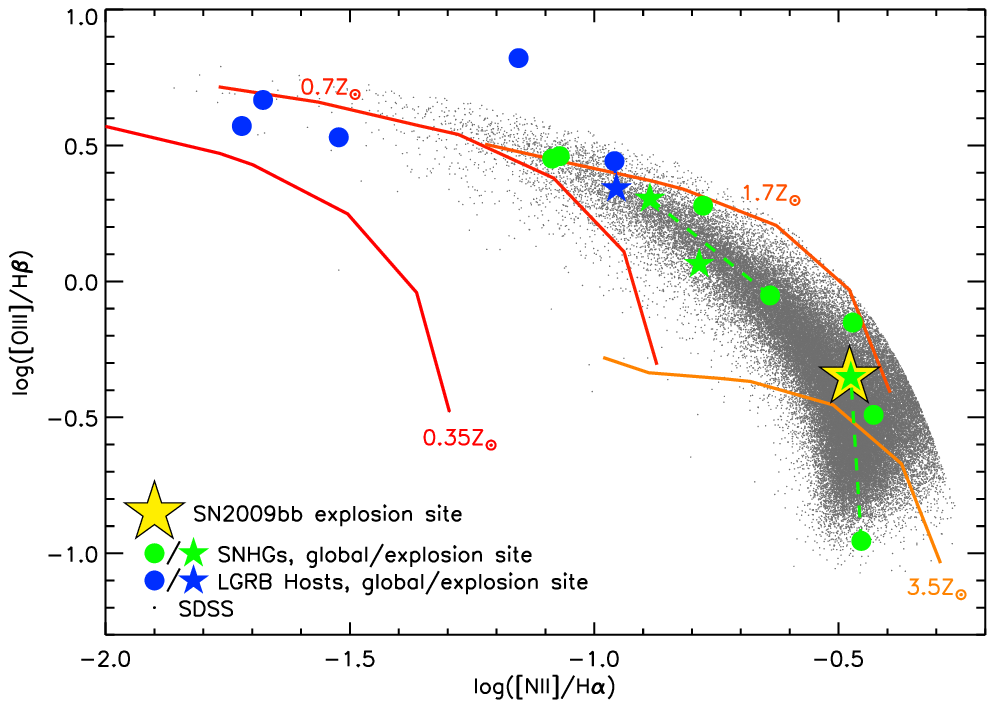}
\plotone{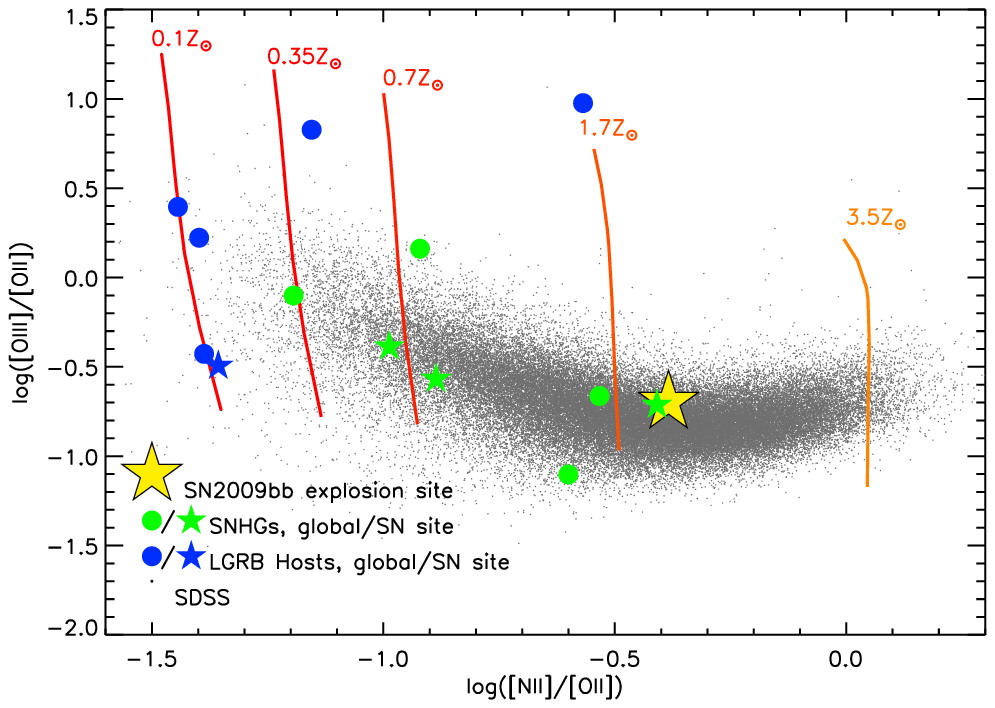}
\caption{Comparison of our SN2009b host galaxy emission line features to a variety of galaxy samples and stellar population synthesis models (Levesque et al.\ (2009b). On the [NII]/H$\alpha$ vs. [OIII]/H$\beta$ (left) and [NII]/[OII] vs. [OIII]/[OII] (right) emission line ratio diagnostic diagrams, we compare the SN2009bb explosion site (yellow star) to ratios for SNe Ic host galaxies and explosion sites (Modjaz et al.\ 2008; green circles and stars), long-duration GRB host galaxies and explosion sites (Levesque et al.\ 2009a, Christensen et al.\ 2008; blue circles and stars), and a sample of 60920 star-forming SDSS galaxies (Kewley et al.\ 2006; points). The Levesque et al.\ (2009b) models assume a zero-age instantaneous burst star formation history, an electron density of $n_e = 100$ cm $^{-3}$, and a range of ionization parameter $-3.5 < \mathcal{U} < -1.9$ (Levesque et al.\ 2009b).}
\end{figure}

\begin{deluxetable}{c c c c c c c c c c c c c}
\tabletypesize{\scriptsize}
\tablewidth{0pc}
\tablenum{1}
\tablecolumns{13}
\tablecaption{\label{tab:gals} SN2009bb Host Environment ISM Properties}
\tablehead{
\multicolumn{8}{c}{Measured Emission Line Fluxes\tablenotemark{a}}
&\colhead{}
&\multicolumn{4}{c}{Derived ISM Properties} \\ \cline{1-8} \cline{10-13}
\colhead{[OII]}
&\colhead{H$\delta$}
&\colhead{H$\gamma$}
&\colhead{H$\beta$}
&\colhead{[OIII]}
&\colhead{[OIII]}
&\colhead{H$\alpha$}
&\colhead{[NII]}
&\colhead{}
&\colhead{E($B-V$)\tablenotemark{b}}
&\colhead{log(O/H) + 12}
&\colhead{Age}
&\colhead{SFR} \\ 
\colhead{3727\AA}
&\colhead{}
&\colhead{}
&\colhead{}
&\colhead{4959\AA}
&\colhead{5007\AA}
&\colhead{}
&\colhead{6584\AA}
&\colhead{}
&\colhead{(mag)}
&\colhead{([NII]/[OII])}
&\colhead{(Myr)}
&\colhead{(M$_{\odot}$ yr$^{-1}$)}
}
\startdata
3.73 &0.88 &1.27 &2.65 &0.39 &1.26 &12.3 &4.09 & &0.48 &8.96 $\pm$ 0.1 &4.5 $\pm$ 0.5 &0.003 \\
\enddata	      	
\tablenotetext{a}{Uncorrected fluxes in units of 10$^{-16}$ ergs cm$^2$ s$^{-1}$ \AA$^{-1}$.}
\tablenotetext{b}{Total color excess in the direction of the galaxy, used to correct for the effects of both Galactic and intrinsic extinction.}
\end{deluxetable}


\begin{references} 
\reference {} Asplund, M., Grevasse, N., \& Sauval, A. J. 2005, in ASP Conf. Ser. Vol. 336, Cosmic Abundances as Records of Stellar Evolution and Nucleosynthesis, ed. T. G. Barnes III \& F. N. Bash, 25
\reference {} Baldwin, J. A., Phillips, M. M., \& Terlevich, R. 1981, Pub. A. S. P., 93, 5
\reference {} Cardelli, J. A., Clayton, G. C., \& Mathis, J. S. 1989, ApJ, 345, 245
\reference {} Christensen, L., Vreeswijk, P. M., Sollerman, J., Th\"{o}ne, C. C., Le Floc'h, E., \& Wiersema, K. 2008, A\&A, 490, 45
\reference {} Copetti, M. V. F., Pastoriza, M. G., \& Dottori, H. A. 1986, A\&A, 156, 111
\reference {} Dessart, L., Burrows, A., Livne, E., \& Ott, C. D. 2008, ApJ, 673, 43
\reference {} Dopita, M. A., Kewley, L. J., Heisler, C. A., \& Sutherland, R. S. 2000, ApJ, 542, 224
\reference {} Foley, R. J. et al.\ 2009, AJ, 138, 376
\reference {} Guetta, D. \& Della Valle, M. 2007, ApJL, 657, L73
\reference {} Heise, J., in't Zand, J., Kippen, R. M., \& Woods, P. M. 2001, in Gamma-Ray Bursts in the Afterglow Era: Proceedings of the International Workshop, vol.
\reference {} Hunter, I. et al.\ 2007, A\&A, 466, 277
\reference {} Kaiser, N. et al. 2002, in Society of Photo-Optical Instrumentation Engineers (SPIE)
Conference Series, Vol. 4836, Society of Photo-Optical Instrumentation Engineers
(SPIE) Conference Series, ed. J. A. Tyson \& S. Wolff, 154Ð164
\reference {} Kelly, P. L., Kirshner, R. P., \& Pahre, M. 2008, ApJ, 687, 1201
\reference {} Kelson, D. D. 2003, PASP, 115, 688
\reference {} Kennicutt, R. C. 1998, ARAA, 36, 189
\reference {} Kewley, L. J. \& Dopita, M. A. 2002, ApJ, 142, 35
\reference {} Kewley, L. J., Groves, B., Kauffmann, G., \& Heckman, T. 2006, MNRAS, 372, 961
\reference {} Kewley, L. J. \& Ellison. S. L. 2008, 681, 1183
\reference {} Kocevski, D., West, A. A., \& Modjaz, M. 2009, ApJ, submitted, arXiv:0905.1953
\reference {} Kulkarni, S. R. et al.\ 1998, Nature, 395, 663
\reference {} Lauberts, A. \& Valentijn, E. A. 1989, The Messenger, 56, 31
\reference {} Law, N. M. et al.\ 2009, PASP, submitted
\reference {} Levesque, E. M., Berger, E., Kewley, L. J., \& Bagley, M. M. 2009a, AJ, submitted
\reference {} Levesque, E. M., Kewley, L. J., \& Larson, K. L. 2009b, AJ, submitted
\reference {} MacFayden, A. I., Woosley, S. E., \& Heger, A. 2001, ApJ, 550, 410
\reference {} Marshall, J. L. et al.\ 2008, SPIE, 7014, 169
\reference {} Massey, P. 2003, ARA\&A, 41, 15
\reference {} Mauch, T. \& Sadler, E. M. 2007, MNRAS, 375, 931
\reference {} Modjaz, M., Kewley, L. J., Kirshner, R. P., Stanek, K. Z., Challis, P., Garnavich, P. M., Greene, J. E., Kelly, P. L., \& Prieto, J. L. 2008, AJ, 135, 1136
\reference {} Osterbrock, D. 1989, Astrophysics of gaseous nebulae and active galactic nuclei (University Science Books)
\reference {} Pignata, G. et al.\ 2009, CBET 1731
\reference {} Piran, T. 1999, Phys. Rep., 314, 575
\reference {} Prieto, J. L., Stanek, K. Z., \& Beacom, J. F. 2008, ApJ, 673, 999
\reference {} Sanders, D. B., Mazzarella, J. M., Kim, D.-C., Surace, J. A., \& Soifer, B. T. 2003, AJ, 126, 1607
\reference {} Schaerer, D., Charbonnel, C., Meynet, G., Maeder, A., \& Schaller, G. 1993, A\&AS, 102, 339
\reference {} Schaerer, D. \& Vacca, W. D. 1998, ApJ, 497, 618
\reference {} Schlegel, D. J., Finkbeiner, D. P., Davis, M. 1998, ApJ, 500, 525
\reference {} Silva, L., Granato, G. L., Bressan, A., \& Danese, L. 1998, ApJ, 509, 103
\reference {} Sim\'{o}n-D\'{i}az, S., Herrero, A., Esteban, C., \& Najarro, F. 2006, A\&A, 448, 351
\reference {} Soderberg, A. M., Nakar, E., Berger, E., \& Kulkarni, S. R. 2006a, ApJ, 638, 930
\reference {} Soderberg, A. M. et al.\ 2006b, Nature, 442, 1014
\reference {} Soderberg, A. M. et al.\ 2009, Nature, submitted
\reference {} Stanek et al.\ 2006, Acta Astron., 56, 333
\reference {} Stritzinger, M., Philips, M. M., Morrell, N., Salgado, F., \& Folatelli, G. 2009, CBET 1751
\reference {} Vink, J. S. \& de Koter, A. 2005, A\&A, 442, 587
\reference {} Wolf, C. \& Podsiadlowski, P. 2007, MNRAS, 375, 1049
\reference {} Woosley, S. E. \& Bloom, J. S. 2006, ARA\&A, 44, 507
\reference {} Woosley, S. E., Heger, A., \& Weaver, T. A. 2002, RMP, 74, 1015
\reference {} Woosley, S. E. \& Heger, A. 2006, ApJ, 637, 914
\reference {} Yoon, S.-C., Langer, N., \& Norman, C. 2006, Astron. Ap. 460, 199
\reference {} Yun, M. S. \& Carilli, C. L. 2002, ApJ, 568, 88
\end{references}
\end{document}